\documentclass[%
 preprint,
 nofootinbib,
 amsmath,amssymb,
 aps,
]{revtex4-2}

\usepackage{macros}
\usepackage{color}
\usepackage{upgreek}
\usepackage{graphicx}
\usepackage{dcolumn}
\usepackage{bm}

\newcommand{\rcoord}{\ensuremath{r}}

\newcommand{\tor}{\zeta}
\newcommand{\phaseFrac}{{\mathit{\Gamma}}}
\newcommand{\adiabat}{\chi}
\newcommand{\adiabattwo}{\mathcal{C}}
\newcommand{\red}[1]{{#1}}
\newcommand{\dblebrck}[1]{[#1]}

\begin{document}

\preprint{APS/123-QED}

\title{A phase-shift-periodic parallel boundary condition for low-magnetic-shear scenarios}

\author{D. A. St-Onge}%
\email[Corresponding e-mail  ]{denis.st-onge@physics.ox.ac.uk}
\author{M. Barnes}%
\affiliation{%
Rudolf Peierls Centre for Theoretical Physics, University of Oxford, Oxford OX1 3PU, United Kingdom 
}%
\altaffiliation[Also at ]{University College, Oxford OX1 4BH, United Kingdom}

\author{F. I. Parra}
\affiliation{
Princeton Plasma Physics Laboratory, Princeton University, Princeton, New Jersey 08544
}%


\date{\today}

\begin{abstract}

We formulate a generalized periodic boundary condition  as a limit of the standard twist-and-shift parallel boundary condition that is suitable for simulations of plasmas with low magnetic shear. This is done by applying a phase shift in the binormal direction when crossing the parallel boundary. While this phase shift can be set to zero without loss of generality in the local flux-tube limit when employing the twist-and-shift boundary condition, we show that this is not the most general case when employing periodic parallel boundaries, and may not even be the most desirable. A non-zero phase shift can be used to avoid the convective cells that plague simulations of the three-dimensional Hasegawa-Wakatani system, and is shown to have measurable effects in periodic low-magnetic-shear gyrokinetic simulations. We propose a numerical program where a sampling of periodic simulations at random pseudo-irrational flux surfaces are used to determine physical observables in a statistical sense. This approach can serve as an alternative to applying the twist-and-shift boundary condition to low-magnetic-shear scenarios which, while more straightforward, can be computationally demanding.

\end{abstract}

\maketitle


\section{\label{sec:intro}Introduction}

Local flux-tube gyrokinetics has been the workhorse of the plasma physics community for the last two decades. By considering an infinitesimally small segment of the device volume, Fourier decomposition can be performed in the directions perpendicular to the magnetic field allowing for fast and efficient numerical simulations with spectral accuracy. In addition, by employing the twist-and-shift boundary condition~\citep{Beer95}, along the magnetic field, the maximal use of the along-the-field resolution can be achieved. This has enabled detailed studies of plasma microturbulence using modest numerical resources. 

The twist-and-shift parallel boundary condition remains the standard for gyrokinetic simulation of both toroidal and non-axisymmetric systems with finite magnetic shear; e.g., $\hat{s} \doteq (\od q/\od r)(r/q) \sim \mathcal{O}(1)$, where $q$ is the magnetic safety factor and $r$ is the minor radial position. Extensions to the twist-and-shift boundary condition have also been proposed for cases with strong magnetic shear, where alternative coordinates or coordinate remappings are used to shift the emphasis away from strongly sheared modes (those with large radial wavenumbers) to those that are unsheared ~\citep{Scott_shiftedmetric, Watanabe_train, Ball_nontwist}. Despite the success of the twist-and-shift boundary condition and other similar methods, scenarios with vanishingly small magnetic shear pose a significant numerical challenge: if certain discreteness conditions are to be met, then the aspect ratio of the simulation $\ell_r / \ell_\alpha$, where $\ell_r$ and $\ell_\alpha$ are the radial and binormal extent of the simulation domain, must scale as $1/N\hat{s}$, where $N$ is the number of poloidal turns, and so such cases require either a large number of poloidal turns or  large radial box sizes. A generalization of the twist-and-shift boundary condition for non-axisymmetric configurations has been recently proposed by~\citet{Martin_2018}, though this approach relies on the local, rather than global, magnetic shear. Many numerical codes in the low-magnetic-shear limit opt instead to use periodic boundary conditions along the magnetic field~\citep{Barnes_stella,gs2}, and indeed some numerical studies have utilized this approach~\citep{Xanthopoulos_2007_nonlin, Faber_2018}.
However, we show in this manuscript that this is not the most general approach for the low-magnetic-shear limit, and  may not be the most desirable one.

This manuscript is organized as follows: in \S \ref{sec:rev}, we give a short review of the twist-and-shift boundary condition, and then reconsider its application to the low-magnetic-shear scenario in \S \ref{sec:lowshear}, formulating a generalized periodic boundary condition for the low-shear limit. We  use the three-dimensional Hasegawa-Wakatani system in \S \ref{sec:HW} to show that using the generalized periodic condition can have significant effects on the system's underlying physical behaviour. In \S \ref{sec:CBC} we use gyrokinetic simulations of the Cyclone Base Case to illustrate how the twist-and-shift boundary condition results in the appearance of mode rational surfaces across the radial simulation domain, and how the generalized periodic boundary can be used to simulate a specific radial location.  Finally, we offer some closing thoughts in \S \ref{sec:outro}.

\section{\label{sec:rev}Review of the twist-and-shift boundary condition}

For local simulations of toroidal plasmas with magnetic shear, the underlying equations (e.g., the gyrokinetic equations) are generally formulated using a field-aligned coordinate system $(\psi, \alpha, \vartheta)$, where the poloidal flux function $\psi$ acts as a radial coordinate, $\vartheta$ is a straight-field line coordinate which denotes the position along a magnetic field line,  $\alpha  = \zeta - q(\psi) \vartheta$ labels the magnetic field line, and $\zeta$ is the toroidal angle. The twist-and-shift parallel boundary condition, which is for the $\vartheta$ dimension, asserts $2\upi N$ periodicity of a quantity $A$ at fixed toroidal angle $\tor$, rather than at fixed $\alpha$,
\begin{equation}\label{eqn:parBC}
    A(\psi, \alpha(\psi,\tor, \vartheta = 0), \vartheta = 0) = A(\psi,\alpha(\psi,\tor, \vartheta =2 \upi N ), \vartheta = 2 \upi N),
\end{equation}
where $N$ is an integer indicating the number of poloidal turns that is chosen as an input parameter for the simulation. In a local flux-tube simulation, the underlying equations are Fourier analysed in the $\psi$ and $\alpha$ dimensions, and thus we can Fourier transform~\eqref{eqn:parBC} and Taylor expand $q(\psi)$, resulting in 
\begin{align}\label{eqn:twistandshift}
   \sum_{k_\psi, k_\alpha} A_\bb{k}(\vartheta = 0) &\rme^{\imag k_\psi (\psi-\psi_0) + \imag k_\alpha (\alpha - \alpha_0)}  \nonumber \\ 
   &\qquad = \sum_{k_\psi, k_\alpha} A_\bb{k}(\vartheta= 2\upi)\rme^{\imag k_\psi (\psi-\psi_0) + \imag k_\alpha (\alpha - \alpha_0)  - 2\upi N \imag k_\alpha [q_0 +  q'(\psi-\psi_0) + \ldots]}.
\end{align}
Here, the subscript zero denotes values taken at the center of the considered domain, $k_\psi$ and $k_\alpha$ are the radial and binormal wavenumbers, and primes denote differentiation with respect to $\psi$. We then further utilize the local flux-tube limit by only keeping the first two terms in the Taylor expansion of $q(\psi) \approx q_0 +  q' (\psi-\psi_0)$, and so the parallel boundary condition amounts to matching a quantity at either end of the $\vartheta$ domain at different radial wavenumber $k_\psi$,
\begin{equation}\label{eqn:twistshift_psi}
A_{k_\psi, k_\alpha}(\vartheta = 0) = C_k A_{k_\psi + \Delta k_\psi,  k_\alpha}(\vartheta= 2\upi), 
\end{equation}
where $C_k = \exp(\imag \theta^{k_\alpha}_\mr{twist} )$ is the phase shift factor, $\theta^{k_\alpha}_\mr{twist} = -2\upi N  k_\alpha  q_0 $ is the phase shift, and $\Delta k_\psi = 2\upi N k_\alpha q'$ is the radial wavenumber connection spacing. Equation~\eqref{eqn:twistshift_psi} allows for an efficient and straightforward implicit treatment of the parallel streaming term, which otherwise can set a stringent constraint on the simulation time step~\citep{gs2,Barnes_stella}. A more comprehensive treatise on the twist-and-shift boundary condition is given by~\citet{Beer95}.

For local flux-tube gyrokinetic simulations, the perpendicular coordinates $\psi$ and $\alpha$ are typically rescaled so that they have units of distance. To wit, we define radial and binormal coordinates $x$ and $y$ given by
\begin{subequations}
\begin{align}
  \label{eqn:xdef}  x &= \frac{q_0}{r_0 B_\mr{ref}}(\psi-\psi_0),\\
  \label{eqn:ydef}  y &= \frac{1}{B_\mr{ref}}\left.\frac{\od \psi}{\od \rcoord}\right|_{\rcoord = \rcoord_0}(\alpha-\alpha_0),
\end{align}
\end{subequations}
where $B_\mr{ref}$ is a reference value of the magnetic field strength and $r_0$ is the location of the flux surface at which all equilibrium and geometric quantities are calculated. Using these normalizations, the connection spacing becomes 
\begin{equation}\label{eqn:connection_spacing}
   \frac{ \Delta k_x}{ k_y} = 2 \upi N \hat{s}.
\end{equation}
This equation can be seen to impose a constraint on the aspect ratio of a numerical simulation: if $\Delta k_x / k_{x0}$ is to be an integer, where $k_{x0} = 2 \upi / \ell_x$ is the smallest resolved finite radial wavenumber and $\ell_x$ is the radial extent of the simulation, then~\eqref{eqn:connection_spacing} leads to the constraint
\begin{equation}\label{eqn:aspect_ratio}
\frac{\ell_x}{\ell_y} = \frac{j_\mr{twist}}{2\pi N \hat{s}},
\end{equation}
where $\ell_y$ is the binormal extent of the simulation and $j_\mr{twist}$ is an adjustable integer parameter that sets the number of distinct radial modes in ballooning space at $k_{y0} = 2 \upi / \ell_y$, the smallest resolved binormal wavenumber. Finally, normalizing the spatial dimensions perpendicular to the magnetic field by the ion thermal gyroradius $\rho_\mr{thi} = v_{\mr{thi}}/\mathit{\Omega}_\mr{i}$, the phase shift becomes 
\begin{equation}\label{eqn:phase_shift}
 \theta^{k_y}_\mr{twist}= -\frac{2 \upi N q_0}{ \rho_\ast} k_y \rho_\mr{thi}  \left(\frac{1}{aB_\mr{ref}}\frac{\od \psi}{\od \rcoord}\right),
\end{equation}
where $a$ is the device minor radius, $\rho_\ast = \rho_\mr{thi} / a$, $v_{\mr{thi}} = \sqrt{2T_\mr{i}/m_\mr{i}}$ is the ion thermal velocity, and $\mathit{\Omega}_\mr{i}$, $T_\mr{i}$ and $m_\mr{i}$ are respectively the ion gyrofrequency, temperature and mass. Bearing in mind that  $ k_y \rho_\mr{thi}  [(\od \psi /\od \rcoord)/a B_\mr{ref}] \sim 1$, \eqref{eqn:phase_shift} reveals that the phase shift is of order $\rho_\ast^{-1}$, and is thus infinitely large in local simulations. While $\theta^{k_y}_\mr{twist}$ can be arbitrarily large, we are free to take its value modulo $2\upi$ and define a new parameter $\phaseFrac$, which we refer to as the phase angle fraction:
\begin{equation}
    \phaseFrac \doteq \frac{2 \upi N q_0}{\rho_\ast} \frac{\rho_\mr{thi}}{\ell_y} \frac{1}{aB_\mr{ref}}\frac{\od \psi}{\od \rcoord} -  \left\lfloor \frac{2 \upi N q_0}{\rho_\ast} \frac{\rho_\mr{thi}}{\ell_y} \frac{1}{aB_\mr{ref}}\frac{\od \psi}{\od \rcoord}\right\rfloor, 
\end{equation}
where $\lfloor A \rfloor$ denotes the floor function. This leads to
\begin{equation}
    C_k = \exp \left(- \imag \phaseFrac k_y \ell_y \right).
\end{equation}
The $\rho_\ast$-large phase shift is thus subsumed in the arbitrary parameter $\phaseFrac$ which falls in the range $[0, 1)$.

In the standard approach to dealing with the twist-and-shift boundary condition for gyrokinetic simulations, the phase shift factor $C_k$ can be set, without loss of generality, to unity. To better understand why this is, it is best to consider the role $j_\mr{twist}$ plays, along with~\eqref{eqn:aspect_ratio}, in determining the simulation domain. The twist-and-shift boundary condition, by including radial magnetic shear,  introduces mode rational surfaces that resonate with $k_{y0}$ at evenly spaced radial locations, the number of which is set by $(k_y /k_{y0})j_\mr{twist}$.
The constraint given by~\eqref{eqn:aspect_ratio} can now be simply understood: in order to maintain radial periodicity, the radial extent must be so that it can be evenly divided by the spacing between adjacent $k_{y0}$ mode rational surfaces, 
\begin{equation}
\Delta x^{k_{y0}}_\mathrm{MRS} = \frac{\ell_y}{  2\upi N \hat{s}},
\end{equation}
with the general spacing for any $k_y$ given by $\Delta x_\mathrm{MRS} = \Delta x^{k_{y0}}_\mathrm{MRS} (k_{y0}/k_y)$. The location of one of the rational flux surfaces that is mode rational for all $k_y$, can found by finding where the term in the square brackets of~\eqref{eqn:twistandshift} vanishes; this is the radial location at which a magnetic field line wraps around on itself, and thus `bites its own tail'. The
condition we must solve for is $q_0 + q'(\psi - \psi_0) = 0$, leading to the simple expression
\begin{equation}
    x_\mr{rational}  = \frac{\ell_x}{j_\mr{twist}}\frac{\theta^{k_{y0}}_\mr{twist}}{2 \upi},
\end{equation}
where we have used the definition of $x$ in~\eqref{eqn:xdef}.\footnote{The radial location of any mode rational surface can be found by solving $Nk_\alpha[q_0 + q'(\psi - \psi_0)] = M$ for $\psi - \psi_0$ where $M \in \mathbb{Z}$. In reduced units this gives $x_\mr{MRS}  =  M\Delta x_\mr{MRS} + x_\mr{rational}$.
}
Thus, the phase shift set by $C_k$ simply translates the location of these surfaces in a radially periodic way~\cite{Ball_prl,Ajay2020}; as all else in the local gyrokinetic model is radially homogeneous, $C_k$ must not have any statistical effect on the resulting physical observables. ($C_k$ can be made not to have any effect whatsoever, provided the initial conditions are also translated accordingly.) This is also compatible with the gyrokinetic ordering: a $\rho_\ast$-small adjustment in the radial position of the simulation domain---and thus $q_0$---allows for $C_k = 1$, and so the twist-and-shift boundary condition remains general even without a phase shift.

\section{\label{sec:lowshear}Low-shear limit}

The limit of vanishing magnetic shear, which is equivalent to $\hat{s} \rightarrow 0$, poses significant numerical challenges, which are apparent in equation~\eqref{eqn:aspect_ratio}: if one wishes to simulate systems with small $\hat{s}$ using the twist-and-shift boundary condition, then one must either utilize a large number of poloidal turns $N$, or be content with extremely large radial extents. This poses a problem for electrostatic codes that use explicit methods, as the long-wavelength ($k_y = 0$, $k_x = k_{x0}$) pseudo-Alfv\'en wave can impose strict constraints on the size of the time step~\citep{Barnes_stella}.
Even for codes that use implicit methods, the back-substitution used in the standard response matrix approach~\citep{gs2,Barnes_stella} scales as $(N_x N_z /j_\mr{twist})^2$, and so for large values of $N_x$ or $N_z$ this operation can become dominant in terms of computation time. 
  
A possible alternative to using the twist-and-shift boundary condition for low-magnetic-shear scenarios is to instead use periodic boundary conditions, which is normally justified by the fact that $\Delta k_x \rightarrow 0$ as $\hat{s} \rightarrow 0$. Indeed, various local gyrokinetic codes take this limit for sufficiently small $\hat{s}$, and some numerical studies have employed this approach in simulations of stellarators with small magnetic shear~\cite{Faber_2018}. Using periodic boundary conditions in place of the twist-and-shift boundary condition is equivalent to ordering $\hat{s} \sim \rho_\ast$, which is akin to applying the local flux-tube limit to the magnetic shear, and so the gyrokinetic equation becomes entirely homogeneous radially. However, while $\Delta k_x$ vanishes in this limit, in principle $C_k$ remains arbitrary;  To the authors’ knowledge, all current flux-tube gyrokinetic set $C_k = 1$ when using pure  periodicity, though this is not the general case. Taking the flux-tube limit for magnetic shear is equivalent to choosing a specific flux surface and making it infinitely wide radially. The phase shift factor $C_k$ is then the parameter that chooses precisely which flux surface to model, and choosing $C_k = 1$ renders the flux surface rational, i.e., mode rational for all values of $k_y$. This choice may not be the most suitable for the low shear limit, since there are infinitely many more irrational surfaces than rational ones for systems with small but finite magnetic shear. The ordering argument mentioned in the previous section also fails here: while a $\rho_\ast$-small change in $q_0$ is needed to set $C_k = 1$, $\hat{s}$ (and thus $\od q / \od r $) is also of order $\rho_\ast$, and so an $\mathcal{O}(1)$ change to the flux surface position must be made to effect this change. We thus advocate for the use of a periodic boundary condition that retains the phase shift factor $C_k$, 
\begin{equation}\label{eqn:phase_shift_bc}
A_{k_\psi, k_\alpha}(\vartheta = 0) = C_k A_{k_\psi,  k_\alpha}(\vartheta= 2\upi),
\end{equation}
otherwise known as a phase-shift-periodic boundary condition. This boundary condition is not new (see, for instance,~\citep{Baver_2011}); however, its utility for cases with small magnetic shear has largely gone unnoticed.

Implementing a phase-shift-periodic boundary condition can be done mainly in two ways. Firstly, one can straightforwardly apply the phase shift when calculating the parallel derivative, which must be done when points on the finite-differencing stencil fall beyond the parallel boundary. Alternatively, we can also implement the phase-shift-periodic boundary condition using a change of coordinates. We define the new coordinates $x'= x$, $z' = z$, and $y' = y- \phaseFrac z(\ell_y / \ell_z)$. This renders the parallel $z$ direction purely periodic at the expense of introducing binormal variation in the parallel derivatives, viz. $\od / \od z = \od /\od z' + (\phaseFrac \ell_y / \ell_z)\od / \od y'$. (Fourier-transformed quantities must also be multiplied by the phase shift $\exp [\imag k_y z' \phaseFrac(\ell_y/\ell_z)]$.) This approach has the advantage of retaining simple triple periodicity, and so is amenable to a Fourier spectral treatment for slab-like (homogeneous in $z$) geometries. 

In the next two sections, we will see that using a phase-shift-periodic boundary condition can have measurable effects on simulations of turbulent plasmas.  This boundary condition may also be a better approximation than periodic boundary conditions for simulations employing the twist-and-shift boundary condition in cases with little magnetic shear.

\section{\label{sec:HW}Convective Cell Mitigation in  Hasegawa-Wakatani}

To illustrate the utility of the phase-shift-periodic boundary condition, we first consider its effect on the three-dimensional Hasegawa-Wakatani equation (HWE). This relatively simple model is computationally inexpensive, and, as we show, its underlying behaviour can be profoundly affected by the use of a phase-shift-periodic boundary condition. The HWE, which is a two-field system that models density-gradient-driven resistive drift wave turbulence, is given simply by
\begin{subequations}
\begin{align}
    \frac{\partial \zeta}{\partial t} + \{\varphi, \zeta\} &= -\adiabat \frac{\partial^2}{\partial z^2}(\varphi - n) - \mu \nabla_\perp^4 \zeta, \\
    \frac{\partial n}{\partial t} + \{\varphi, n\} &= - \adiabat \frac{\partial^2}{\partial z^2}(\varphi - n) - \kappa \frac{\partial \varphi}{\partial y} - \mu \nabla_\perp^4 n, 
\end{align}
\end{subequations}
where $n$ is the density perturbation, $\varphi$ is the electrostatic potential, $\zeta = \nabla_\perp^2 \varphi$ is the vorticity, $\adiabat$ is the adiabaticity parameter, $\kappa$ parameterizes the background density gradient,  $\mu$ is a diffusion coefficient, and $\{A,B\} = \partial_x A \partial_y B - \partial_y A \partial_x B$ is the Poisson bracket. While this system can be derived in three-dimensional geometry, it is most often studied numerically and analytically in its two-dimensional reduction, which is performed by the substitution $-\adiabat \partial_z^2 \rightarrow \mathcal{C}$, where $\adiabattwo$ is now another tuneable model parameter. Early studies simply made $\adiabattwo$ a constant, but it was later realized that this resulted in artificially large screening of $k_y = k_z = 0$ zonal modes for systems with finite magnetic shear~\citep{Numata}. The remedy was to enforce $k_z = 0$ for $k_y = 0$ modes  by replacing $\adiabattwo $ with the operator $\hat{\adiabattwo}$, given by
\begin{equation}
    \hat{\adiabattwo}A = \adiabattwo \left(A - \frac{1}{\ell_y}\int_0^{\ell_y} \od y \, A\right),
\end{equation}
which subtracts from $A$ its flux-surface-averaged component. Spectrally, this is equivalent to $\hat{\adiabattwo} = \adiabattwo$ for non-zonal modes and zero otherwise. The resulting modification leads to the so-called modified-Hasegawa-Wakatani equation, which has been a paradigmatic model for discussing the interaction between zonal flows and turbulence. 

Currently, a satisfactory comparison of the three-dimensional Hasegawa-Wakatani system to its two dimensional simplifications has yet to be carried out; the need for such a study has now become even more compelling, as an alternative two-dimensional HWE has recently been proposed by~\citet{Majda_2018}. However, a direct comparison between the two- and three- dimensional systems is made difficult by the tendency of the three-dimensional HWE in a triply-periodic domain to condensate the majority of the free energy into $k_z = 0$ convective cells~\citep{Biskamp_1995}, a behaviour not shared with its two-dimensional counterparts. These cells eventually grow to be box-scale, resulting in an almost complete quenching of the underlying drift-wave turbulence. To mitigate these convective cells,~\citet{Korsholm_1999} advocated the use of an \emph{ad hoc} non-periodic radial boundary condition with damping operators at the boundaries, though it has also been shown that effective mitigation can be attained by adopting some amount of magnetic shear~\citep{Hasegawa_Wakatani, Kammel_thesis}. We will show here, however, that the radial inhomogeneity resulting from finite magnetic shear is actually unnecessary; all that is needed is the across-the-boundary parallel variation for $k_y \ne 0$ modes that arises from a phase-shift-periodic boundary condition. This variation has the effect of enforcing $k_y = 0$ for $k_z = 0$ modes, and thus convective cells, which require finite $k_y$, can be avoided. The subsequent evolution of the system then closely resembles the two-dimensional modified HWE, where zonal flows govern the turbulence.

To demonstrate this, we perform pseudospectral simulations of the three-dimensional HWE using two sets of boundary conditions, the first using a triply periodic domain, while the second applies a phase shift across the parallel boundary. The radial and binormal extents are set to $\ell_x = \ell_y = 20 \upi \rho_\mathrm{s} $, \red{where $\rho_\mathrm{s} = \sqrt{T_\mr{e}/m_\mr{i}}$ is the ion sound radius, $T_
\mr{e}$ is the electron temperature and $m_\mr{i}$ is the ion mass}; in principle, the parallel extent can be set to unity with a proper  renormalization of $\adiabat$, though in practice we use $\ell_z = 2 \upi$ to keep in line with the convention of using $z$ as a poloidal-like variable in gyrokinetic simulations. We set $\kappa = 1$,  $\adiabat = 0.125$, and $\mu = 0.01$. Linear terms are handled implicitly using a second-order Crank-Nicholson method,  the nonlinearity is evaluated using the third-order Adams-Bashforth method, and 3/2 padding is used in all dimensions for dealiasing. Finally, simulations are seeded with small-amplitude random noise.

\begin{figure}
    \centering
    \includegraphics[width=\textwidth]{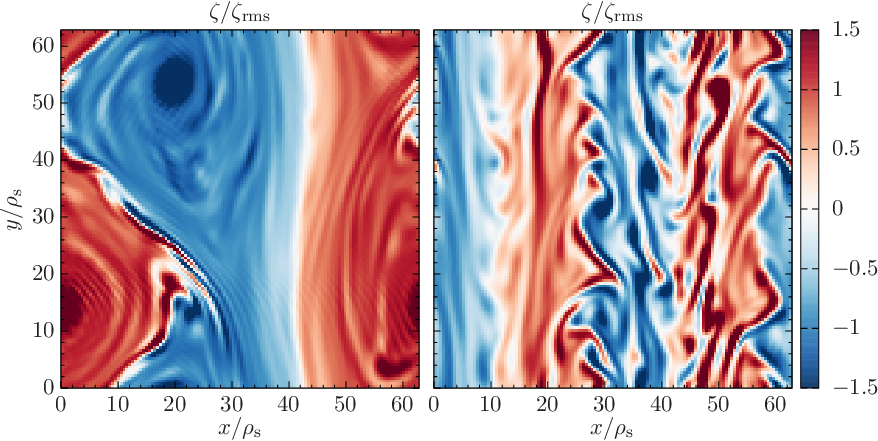}
    \caption{Snapshot of the vorticity, scaled by its root-mean-squared value, at $z = 0$ for simulations employing a periodic boundary conditions without (left)  and with (right) a phase shift across the parallel boundary. Convective cells dominate the purely periodic system, whereas the inclusion of a phase shift results in behaviour that resembles the two-dimensional modified Hasegawa-Wakatani system.}
    \label{fig:vorticity}
\end{figure}

Figure~\ref{fig:vorticity} shows snapshots of the ion vorticity $\zeta$, normalized by its root-mean-squared value $\zeta_\mr{rms}$, at $z = 0$ and late times for the cases with $\phaseFrac = 0$ (left) and $\phaseFrac = 59/128$ (right). While the latter value is not strictly irrational, since $59$ is relatively prime with $128$ (the number of collocation points in the simulation), it is enough to ensure that there are no values of $k_y \ne 0$ for which the surface is precisely mode rational. For the simulation employing purely periodic boundaries ($\phaseFrac =0$), large-scale $k_z = 0$ convective cells develop which dominate the subsequent dynamics of the system. Unlike zonal flows, these modes are able to self-interact and absorb surrounding $k_z \ne 0$ turbulence. On the other hand, for $\phaseFrac = 59/128$, these convective cells do not develop. Instead, $k_z = k_y = 0$ zonal flows quickly emerge instead, which have the effect of regulating the turbulence. The latter scenario is more familiar to those who study gyrokinetics and the two-dimensional mHWE. That the three-dimensional system behaves much like its two-dimensional counterpart shows that comparisons between the two are indeed possible. In fact, a phase-shift-periodic boundary condition makes clear how the mHWE can be obtained from the three-dimensional equations, and so the two-dimensional system is not as \emph{ad hoc} as it first appears. This calls into question whether the modifications proposed by~\citet{Majda_2018} are physical, or even necessary. A more in-depth comparison between the two- and three-dimensional models can now be performed to determine exactly which of the two-dimensional reductions is most viable, though this is beyond the scope of the current manuscript.

\section{\label{sec:CBC}Cyclone Base Case with vanishing shear}

In this section we perform electrostatic gyrokinetic simulations of the CBC~\citep{Dimits}\red{---a highly driven plasma exhibiting strong toroidal ion-temperature-gradient turbulence---}with a much reduced value of normalized magnetic shear ($\hat{s} = 0.01$ versus the nominal $\hat{s} = 0.796$).\footnote{\citet{Volcokas_2022} have found that for CBC, the phase-shift-periodic boundary condition becomes applicable for $\hat{s}\lesssim 0.1$.} The simulations are performed using the gyrokinetic~\citep{FriemanChen,Dubin} flux-tube code~\texttt{stella} in local operation, the details of which are given in~\citet{Barnes_stella}, \red{with additional numerical tests being performed in~\citet{stonge_stella}}. Both the twist-and-shift and phase-shift-periodic boundary conditions are utilized in order to make comparisons. 
All simulations use 41 dealiased binormal modes with a binormal extent of $\ell_y = 20 \upi \rho_\mr{thi}$, 12 points along the direction parallel to the magnetic field, 36 points in the parallel velocity, 8 points for the magnetic moment $\mu$, \red{and a single poloidal turn ($N = 1$ and $l_z = 2\upi$)}. The simulation employing the twist-and-shift boundary condition uses 683 dealiased radial modes with $j_\mr{twist} = 1$, leading to a radial extent of $\ell_x = 1000\rho_\mr{thi}$; the simulations employing the phase-shift-periodic boundary condition use 41 dealiased radial modes with a radial extent $\ell_x = \ell_y = 20 \upi  \rho_\mr{thi}$. Kinetic electrons are used, with a mass ratio given by $m_\mr{i}/ m_\mr{e} = 3672$. \red{As is usual for tokamak turbulence, we assume that the electrostatic potential is determined by the quasineutrality equation, discarding the Debye length scale.} A small amount of like-species collisional damping is provided by a Dougherty collision operator with a collision frequency of $\nu_s (a/v_{\mr{th}s}) = 0.005$. All simulations employ 3/2 padding for dealiasing, and are initialized with low-amplitude random noise, out of which ion-temperature-gradient instabilities develop, saturating nonlinearly.

\begin{figure}
    \centering
    \includegraphics[width=\textwidth]{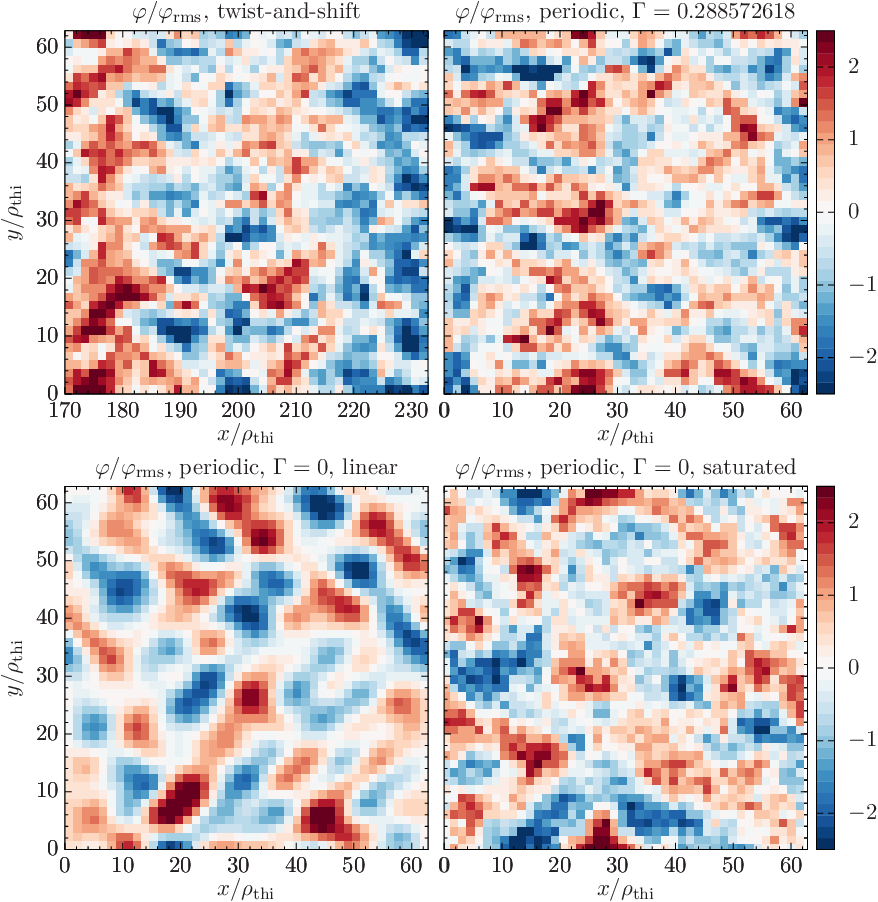}
    \caption{Snapshot of the electrostatic potential, scaled by its root-mean-squared value, at the outboard midplane for gyrokinetic simulations of CBC  employing the twist-and-shift boundary condition (top left)  and with phase shift periodic boundaries with $\Gamma = 0.288572618 $(top right), as well as a purely periodic simulation in the linear phase (bottom left) and saturated state (bottom right). Note that only a subset of the domain in the upper left panel is shown in order to match the radial scale of the other panels. }
    \label{fig:phi_omp}
\end{figure}

\begin{figure}
    \centering
    \includegraphics[width=0.8\textwidth]{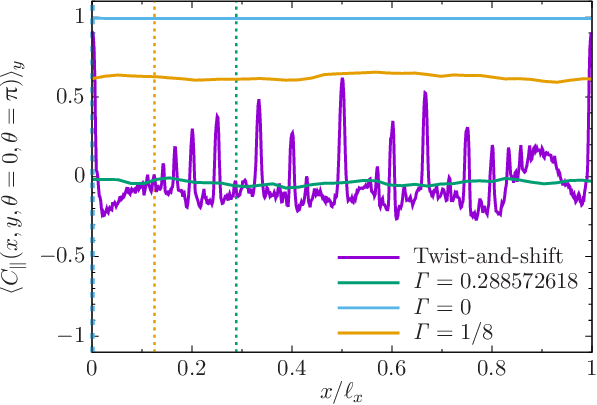}
    \caption{\red{ Parallel correlation function, averaged over the binormal direction, for simulations employing the twist-and-shift boundary condition as well as the phase-shift-periodic boundary condition with various values of $\phaseFrac$. Vertical dotted lines denote the locations at which the periodic simulations are coincident with the simulation employing the twist-and-shift boundary condition.} }
    \label{fig:cpar}
\end{figure}

Snapshots of the electrostatic potential $\varphi$ at the outboard midplane  are shown in figure~\ref{fig:phi_omp} for a selection of the simulations mentioned in the previous paragraph. In general, the saturated states of the simulation employing the twist-and-shift boundary condition (top left) and periodic simulations using an irrational phase angle fraction (top right) bear a strong resemblance, being representative of the typical zonal-flow-mediated strongly driven drift-wave turbulence of the CBC. Even at these modest resolutions, the radially elongated eddies can be observed in these snapshots. On the other hand, simulations employing periodic boundary conditions with low-order rational phase angle fractions---namely, $0$  and $1/2$---exhibit a different type of turbulence which is more homogenous. This may be due to the most unstable modes now having a finite radial wavenumber, which can be seen in the linear state for $\Gamma = 0$ (bottom left); signatures of these oblique modes can still be discerned in the saturated state (bottom right). These structures in the periodic simulations cannot be considered a good approximation to the simulation employing the twist-and-shift boundary condition. Interestingly, these periodic simulations also do not develop the convective cells observed in the HWE, though due to toroidicity these gyrokinetic simulations are not homogeneous in the parallel direction. 

\red{The parallel mode structure in these simulations also differs greatly when the parallel boundary condition is changed. In figure~\ref{fig:cpar}, we compute the quasilinearlation function~\citep{Ball_prl}, 
\begin{equation}\label{eq:cpar}
    C_\parallel(x,y,\theta_1,\theta_2) = \frac{\langle\varphi_\mr{DW}(x,y,\theta_1)\varphi_\mr{DW}(x,y,\theta_2)\rangle_t}{\sqrt{\langle\varphi_\mr{DW}^2(x,y,\theta_1)\rangle_t\langle\varphi_\mr{DW}^2(x,y,\theta_2)\rangle_t}}
\end{equation}
where $\varphi_\mr{DW} =\varphi - \ell_y^{-1}\int \varphi \od y $ is the $k_y =0$ drift-wave component of the electrostatic potential and $\langle \cdots\rangle_t$ denotes time averaging over the saturated state. Spikes in the correlation function near low-order rational surfaces for the simulation employing the twist-and-shift boundary condition indicate locations where self-interaction is prominent; this finding has been previously reported in~\citet{Ball_prl}. Interestingly, the parallel correlation function of the phase-shift-periodic simulations match closely the twist-and-shift simulation at the appropriate locations, i.e., $\Gamma = x/\ell_x$ for $j_\mr{twist} = N = 1$, and thus the phase-shift-periodic boundary condition captures the correct local physics, whether at a rational or irrational surface. }

\begin{figure}
    \centering
    \includegraphics[width=0.8\textwidth]{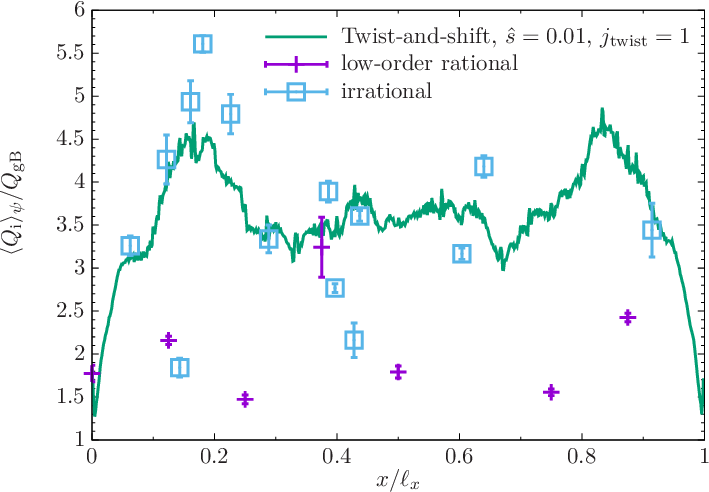}
    \caption{Flux-surface-averaged ion heat flux, normalized by its gyroBohm value $Q_\mr{GB}$, as a function of radius. Solid line denote resulting profile from a simulation, time-averaged over a short window directly after the linear phase ends.
    Crosses denote periodic simulations using low-order rational values for $\phaseFrac$, while squares denote periodic simulations using irrational values of $\phaseFrac$.}
    \label{fig:flux_profile}
\end{figure}

Figure~\ref{fig:flux_profile} displays the radial profile of the flux-surface-averaged ion heat flux $Q_\mr{i}$ for the simulation employing the twist-and-shift boundary condition. This flux, which has been normalized by its gyroBohm value $Q_\mr{gB} = n_\mr{i}T_\mr{i}v_\mr{thi}\rho_\mr{i}^2/R_0^2$ (where $n_\mr{i}$ is the ion density and $R_0$ is the major radius at the center of the flux surface), 
has also been time-averaged over the entire saturated state \dblebrck{$t(v_\mr{thi}/a) > 150$}. 
Included in this plot are the volume-averaged heat fluxes from the periodic simulations employing low-order rational values (crosses) and irrational values (boxes) for $\phaseFrac$. (The exact values of $\phaseFrac$ are given in table~\ref{tab:phases}.) The radial profile of the heat flux for the twist-and-shift simulation reveals a rich structure resulting from the mode rational surfaces introduced by the twist-and-shift boundary condition. Consistent with the findings in~\citet{Ajay2020}, shear layers around the rational surface at $x = 0$ strongly stabilize the turbulence near that region, which is also reflected in the purely periodic simulation (i.e., $\phaseFrac = 0$). \red{Interestingly, while the averaged heat flux profile is mostly flat, it does experience a dip at the rational surface, resulting in an exacerbation of the local temperature gradient.
} 
\red{Surprisingly, the behaviour of the periodic simulations is strongly affected by the value of the phase shift, and the resulting ion heat flux can vary by almost an order of magnitude. \red{This is partially due to a change in the linear growth rates from modifications to the electron response, as well as differences in zonal flow behaviour that result from the stronger self-interaction
in systems that are closer to rational surfaces.}\footnote{ Neoclassical zonal flow physics~\citep{Rosenbluth} is independent of the use of twist-and-shift or phase-shift-periodic boundary conditions, as these modes remain purely periodic, viz. equations~\eqref{eqn:twistandshift} and~\eqref{eqn:phase_shift_bc} with $k_\alpha = 0$.}} \red{Recently,~\citet{Volcokas_2022} observed the appearance of long turbulent eddies in the low-shear CBC whose binormal periodicity are resonant with the order of the flux surface, and thus high-order surfaces feature eddies with large binormal wavenumbers. If one assumes the turbulence is approximately isotropic in the perpendicular plane, quasilinear arguments indicate that the resultant heat flux would be reduced ($Q \sim  \gamma / k_\perp^2$, where $\gamma$ is the growth rate). This is qualitatively consistent with the findings in~\citep{Volcokas_2022} and with the simulation results reported here, where large binormal wavenumbers are witnessed for periodic simulations near rational surfaces.  }
Apart from $\phaseFrac = 0$, the simulations using low-order rational values of $\phaseFrac$ tend to not satisfactorily approximate the local heat flux. The opposite, however, is generally true for the simulations employing irrational values. This is not too surprising, since flux surfaces are densely irrational when the magnetic shear is finite.

\begin{figure}
    \centering
    \includegraphics[width=0.6\textwidth]{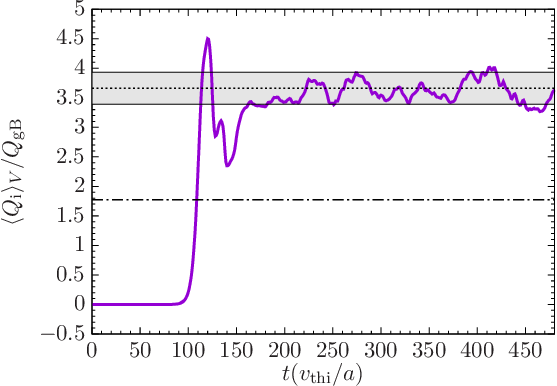}
    \caption{Time evolution of the volume-averaged ion heat flux, normalized to its gyroBohm value, for the simulation using the twist-and-shift boundary condition. Dashed line denotes the average total heat flux of the fourteen periodic simulations employing irrational phase angle fractions, while the shaded region denotes the standard error of that mean.}
    \label{fig:heat_flux}
\end{figure}

The time evolution of the volume-averaged ion heat flux for the simulation using the twist-and-shift boundary condition is shown in figure~\ref{fig:heat_flux}.  We also display the mean volume-averaged heat flux averaged over all fourteen periodic simulations employing irrational values of $\phaseFrac$. It is seen from this figure that the mean of the fourteen periodic simulations approximates the total heat flux of the full twist-and-shift simulation extremely well. Importantly, the computational savings gained by using periodic boundary conditions is enormous: the fourteen periodic simulations used a combined total of 480 CPU hours, while the single twist-and-shift simulation cost 10,000 CPU hours, a roughly factor of 20 savings. Thus, using a phase-shift-periodic boundary condition can be a possible route to cheaper numerical campaigns of  low-magnetic-shear scenarios. Note that while we used fourteen simulations in this manuscript, a much smaller number could have been used to still achieve the correct flux to within a few percent. 

\begin{table}
\begingroup
\setlength{\tabcolsep}{10pt} 
    \centering
    \begin{tabular}{ccccc}
    \hline
    \hline
        $\phaseFrac$ (rational) &  & \multicolumn{3}{c}{$\phaseFrac$ (irrational) } \\
        \hline
        0       && 0.0623970386 &&   0.386142693  \\
        1/8   && 0.121690317 && 0.396917828 \\
        1/4    && 0.14317397 && 0.427936569\\
        3/8     && 0.16086218165 && 0.43719857 \\
        1/2 && 0.180877018  && 0.60393697877 \\
        5/8 && 0.226316958 && 0.639988589286 \\
        7/8 && 0.288572618 && 0.91438395191\\
          \hline
         \hline
    \end{tabular}
    \caption{A list of phase angle fractions $\phaseFrac$ used for the periodic simulations employing low-order rational values (left column) and irrational values (center and right columns).}
    \label{tab:phases}
\endgroup
\end{table}

\section{\label{sec:outro}Discussion}

In this manuscript we demonstrated the utility of the phase-shift-periodic boundary condition to scenarios using vanishingly small values of global magnetic shear. \red{This was done using simulations of the three-dimensional Hasegawa-Wakatani equations, which showed that in certain systems the underlying behaviour can be completely altered by applying a phase shift across the parallel boundary. In addition, gyrokinetic simulations of the Cyclone Base Case with a modified value of the normalized magnetic shear $\hat{s}$ were shown to exhibit a strong dependence on the phase shift across the parallel boundary. In principle, the phase-shift-periodic boundary condition should be generically useful for any system with small-but-finite values of magnetic shear.}

With the usefulness of a phase-shift-periodic boundary condition established, we now envision two possible numerical programs when studying systems in the low-magnetic-shear regime. The first program is the simpler of the two: one performs a single numerical simulation with a very deliberate choice of the phase angle fraction $\phaseFrac$. If, on the other hand, the numerical worker is agnostic to the precise value of the field line pitch that is used, then a second program can be undertaken, wherein a number of phase-shift-periodic flux-tube simulations are performed, each with a random but irrational value of $\phaseFrac$, the number of simulations being determined by the desired tolerance interval on the standard error of the mean. The resulting physical observables can then be averaged accordingly, on which statements and conclusions can then be made. 

In the context of the HWE equation, the phase-shift-periodic boundary condition paves the way for the comparison between its two- and three-dimensional settings. Such a comparison would be helpful in evaluating the usefulness of various two-dimensional reductions of the HWE. The phase-shift-periodic boundary condition could be useful for other fluid and kinetic systems formulated in the low-magnetic-shear limit, such as the three-field ion-temperature-gradient system proposed by~\citet{Ivanov_2022}. Finally, in the context of gyrokinetic simulations of turbulent plasmas, future work should continue to evaluate the use of phase-shift-periodic boundary conditions for more physically relevant situations; while the phase-shift-periodic boundary conditions perform well for the Cyclone Base Case studied here, ideally it should also be tested for cases with more complex flux-surface shaping, as well as for non-axisymmetric geometries (stellarators).

\vspace{0.2cm}

\begin{acknowledgments}

The authors would like to thank Plamen G. Ivanov for many fruitful discussions. This work has been carried out within the framework of the EUROfusion Consortium, funded by the European Union via the Euratom Research and Training Programme (Grant Agreement No 101052200 - EUROfusion). Views and opinions expressed are however those of the author(s) only and do not necessarily reflect those of the European Union or the European Commission. Neither the European Union nor the European Commission can be held responsible for them. This work was supported in part by the Engineering and Physical Sciences Research Council (EPSRC) [Grant Number  EP/R034737/1]. The authors acknowledge EUROfusion, the EUROfusion High Performance Computer (Marconi-Fusion) under the project MULTEI and OXGK, and the use of ARCHER through the Plasma HEC Consortium [EPSRC Grant Number EP/R029148/1] under the project e607. This work was also supported by the U.S. Department of Energy under contract number DE-AC02-09CH11466. The United States Government retains a non-exclusive, paid-up, irrevocable, world-wide license to publish or reproduce the published form of this manuscript, or allow others to do so, for United States Government purposes.

\end{acknowledgments}

\section*{Source code}

The \texttt{MATLAB} script used to perform the simulations of the three-dimensional HWE equations can be obtained at the GitHub repository \url{https://github.com/DenSto/HWE_solver}.
The \texttt{stella} source code is available at the GitHub repository~\url{https://github.com/stellaGK/stella}, while documentation can be found at ~\url{https://stellagk.github.io/stella/}.

\section*{Data availability statement}

{
\sloppy The input files that are used for the gyrokinetic simulations featured in this manuscript, as well as instructions for their use, are available at the following URL: \url{https://doi.org/10.5287/bodleian:ORbV2dNRJ}.
}


\bibliography{refs}

\end{document}